\documentclass[a4paper,10pt]{aastex}
\usepackage{array}
\title{Multiwavelength Observations of LS I +61$^{\circ}$ 303 with VERITAS, Swift and RXTE}

\author{
V. A. Acciari\altaffilmark{1},
E. Aliu\altaffilmark{2},
T. Arlen\altaffilmark{3},
M. Bautista\altaffilmark{4},
M. Beilicke\altaffilmark{5},
W. Benbow\altaffilmark{6},
M. B{\"o}ttcher\altaffilmark{7},
S. M. Bradbury\altaffilmark{8},
V. Bugaev\altaffilmark{5},
Y. Butt\altaffilmark{9},
Y. Butt\altaffilmark{9},
K. Byrum\altaffilmark{10},
A. Cannon\altaffilmark{11},
A. Cesarini\altaffilmark{12},
Y. C. Chow\altaffilmark{3},
L. Ciupik\altaffilmark{13},
P. Cogan\altaffilmark{4},
P. Colin\altaffilmark{14},
W. Cui\altaffilmark{15},
M. Daniel\altaffilmark{8,\dag},
R. Dickherber\altaffilmark{5},
T. Ergin\altaffilmark{9},
A. Falcone\altaffilmark{16},
S. J. Fegan\altaffilmark{3,\dag\dag},
J. P. Finley\altaffilmark{15},
P. Fortin\altaffilmark{17},
L. Fortson\altaffilmark{13},
A. Furniss\altaffilmark{18},
D. Gall\altaffilmark{15},
G. H. Gillanders\altaffilmark{12},
J. Grube\altaffilmark{11},
R. Guenette\altaffilmark{4},
G. Gyuk\altaffilmark{13},
D. Hanna\altaffilmark{4},
E. Hays\altaffilmark{19},
J. Holder\altaffilmark{2},
D. Horan\altaffilmark{20,\dag\dag},
C. M. Hui\altaffilmark{14},
T. B. Humensky\altaffilmark{21},
P. Kaaret\altaffilmark{22},
N. Karlsson\altaffilmark{13},
D. Kieda\altaffilmark{14},
J. Kildea\altaffilmark{6},
A. Konopelko\altaffilmark{23},
H. Krawczynski\altaffilmark{5},
F. Krennrich\altaffilmark{24},
M. J. Lang\altaffilmark{12},
S. LeBohec\altaffilmark{14},
G. Maier\altaffilmark{4},
A. McCann\altaffilmark{4},
M. McCutcheon\altaffilmark{4},
J. Millis\altaffilmark{25},
P. Moriarty\altaffilmark{1},
R. Mukherjee\altaffilmark{17},
T. Nagai\altaffilmark{24},
R. A. Ong\altaffilmark{3},
A. N. Otte\altaffilmark{18},
D. Pandel\altaffilmark{22},
J. S. Perkins\altaffilmark{6},
J. S. Perkins\altaffilmark{6},
M. Pohl\altaffilmark{24},
J. Quinn\altaffilmark{11},
K. Ragan\altaffilmark{4},
L. C. Reyes\altaffilmark{26},
P. T. Reynolds\altaffilmark{27},
E. Roache\altaffilmark{6},
H. Joachim Rose\altaffilmark{8},
M. Schroedter\altaffilmark{24},
G. H. Sembroski\altaffilmark{15},
A. W. Smith\altaffilmark{10,*},
D. Steele\altaffilmark{13},
M. Stroh\altaffilmark{16},
S. Swordy\altaffilmark{21},
M. Theiling\altaffilmark{6},
J. A. Toner\altaffilmark{12},
A. Varlotta\altaffilmark{15},
V. V. Vassiliev\altaffilmark{3},
R. G. Wagner\altaffilmark{10},
S. P. Wakely\altaffilmark{21},
J. E. Ward\altaffilmark{11},
T. C. Weekes\altaffilmark{6},
A. Weinstein\altaffilmark{3},
R. J. White\altaffilmark{8},
D. A. Williams\altaffilmark{18},
S. Wissel\altaffilmark{21},
M. Wood\altaffilmark{3},
B. Zitzer\altaffilmark{15}
}
\altaffiltext{*}{Corresponding Author: awsmith@hep.anl.gov}
\altaffiltext{1}{Department of Life and Physical Sciences, Galway-Mayo Institute of Technology, Dublin Road, Galway, Ireland}
\altaffiltext{2}{Department of Physics and Astronomy and the Bartol Research Institute, University of Delaware, Newark, DE 19716, USA}
\altaffiltext{3}{Department of Physics and Astronomy, University of California, Los Angeles, CA 90095, USA}
\altaffiltext{4}{Physics Department, McGill University, Montreal, QC H3A 2T8, Canada}
\altaffiltext{5}{Department of Physics, Washington University, St. Louis, MO 63130, USA}
\altaffiltext{6}{Fred Lawrence Whipple Observatory, Harvard-Smithsonian Center for Astrophysics, Amado, AZ 85645, USA}
\altaffiltext{7}{Astrophysical Institute, Department of Physics and Astronomy, Ohio University, Athens, OH 45701}
\altaffiltext{8}{School of Physics and Astronomy, University of Leeds, Leeds, LS2 9JT, UK}
\altaffiltext{9}{Harvard-Smithsonian Center for Astrophysics, 60 Garden Street, Cambridge, MA 02138, USA}
\altaffiltext{10}{Argonne National Laboratory, 9700 S. Cass Avenue, Argonne, IL 60439, USA}
\altaffiltext{11}{School of Physics, University College Dublin, Belfield, Dublin 4, Ireland}
\altaffiltext{12}{School of Physics, National University of Ireland, Galway, Ireland}
\altaffiltext{13}{Astronomy Department, Adler Planetarium and Astronomy Museum, Chicago, IL 60605, USA}
\altaffiltext{14}{Physics Department, University of Utah, Salt Lake City, UT 84112, USA}
\altaffiltext{15}{Department of Physics, Purdue University, West Lafayette, IN 47907, USA }
\altaffiltext{16}{Department of Astronomy and Astrophysics, 525 Davey Lab, Pennsylvania State University, University Park, PA 16802, USA}
\altaffiltext{18}{Santa Cruz Institute for Particle Physics and Department of Physics, University of California, Santa Cruz, CA 95064, USA}
\altaffiltext{19}{N.A.S.A./Goddard Space-Flight Center, Code 661, Greenbelt, MD 20771, USA}
\altaffiltext{20}{Laboratoire Leprince-Ringuet, Ecole Polytechnique, CNRS/IN2P3, F-91128 Palaiseau, France}
\altaffiltext{21}{Enrico Fermi Institute, University of Chicago, Chicago, IL 60637, USA}
\altaffiltext{22}{Department of Physics and Astronomy, University of Iowa, Van Allen Hall, Iowa City, IA 52242, USA}
\altaffiltext{23}{Department of Physics, Pittsburg State University, 1701 South Broadway, Pittsburg, KS 66762, USA}
\altaffiltext{24}{Department of Physics and Astronomy, Iowa State University, Ames, IA 50011, USA}
\altaffiltext{25}{Department of Physics, Anderson University, 1100 East 5th Street, Anderson, IN 46012}
\altaffiltext{26}{Kavli Institute for Cosmological Physics, University of Chicago, Chicago, IL 60637, USA}
\altaffiltext{27}{Department of Applied Physics and Instumentation, Cork Institute of Technology, Bishopstown, Cork, Ireland}

\altaffiltext{\dag}{Currently at University of Durham, Department of Physics, South Road, Durham DH1 3LE, U.K.} 
\altaffiltext{\dag\dag}{Currently at Laboratoire Leprince-Ringuet, Ecole Polytechnique, CNRS/IN2P3, F-91128 Palaiseau, France}

\begin{document}
\begin{abstract}

We present results from a long-term monitoring campaign on the TeV binary LSI +61$^{\circ}$ 303 with VERITAS at energies above 500 GeV, and in the 2-10 keV hard X-ray bands with RXTE and Swift, sampling nine 26.5 day orbital cycles between September 2006 and February 2008. The binary was observed by VERITAS to be variable, with all integrated observations resulting in a detection at the 8.8$\sigma$ (2006/2007) and 7.3$\sigma$ (2007/2008) significance level for emission above 500 GeV. The source was detected during active periods with flux values ranging from 5 to 20$\%$ of the Crab Nebula, varying over the course of a single orbital cycle. Additionally, the observations conducted in the 2007-2008 observing season show marginal evidence (at the 3.6$\sigma$ significance level) for TeV emission outside of the apastron passage of the compact object around the Be star. Contemporaneous hard X-ray observations with RXTE and Swift show large variability with flux values typically varying between 0.5 and 3.0 $\times10^{-11}$ ergs cm$^{-2}$ s$^{-1}$ over a single orbital cycle. The contemporaneous X-ray and TeV data are examined and it is shown that the TeV sampling is not dense enough to detect a correlation between the two bands.

\end{abstract}
\newpage
\pagebreak

\pagebreak

\section{Introduction}

LS I +61$^{\circ}$ 303 is one of the most extensively studied binary star systems in the Milky Way and, although it has been the subject of many observational campaigns, the true nature (i.e. microquasar or binary pulsar) of the system remains unclear. The system can be classified as a high mass X-ray binary (HMXB) located at a distance of $\sim$2 kpc; the components of the system consisting of a compact object in a 26.496 ($\pm$0.003) day orbit around a massive BO Ve main sequence star (Hutchings and Crampton 1981, Casares et al. 2005). The motion of the compact object around its main sequence companion is traditionally characterized by the orbital phase, $\phi$, ranging from 0.0 to 1.0. $\phi=0$ is set at  JD 2443366.775 (Gregory and Taylor 1978), with periastron passage believed to occur at $\phi$=0.23$\pm0.02$ (Casares et al. 2005) or $\phi$=0.30$\pm0.01$ (Grundstrom et al. 2006), and apastron passage between $\phi$=0.65 and $\phi$=0.85. Historically, LS I +61$^{\circ}$ 303 has been an object of interest due to its periodic outbursts at radio (Paredes et al. 1998, Gregory 2002) and  X-ray energies (Leahy et al. 1997, Taylor et al. 1996, Greiner and Rau 2001, Harrison et al. 2000). The radio outbursts are well correlated with the orbital phase (Gregory 2002), although the phase of maximum emission can vary between $\phi$=0.45 and $\phi$=0.95.  LS I +61$^{\circ}$ 303 was first identified at gamma-ray energies with the COS-B source 2CG 135 +01 (Hermsen et al. 1977) and has also been identified with the EGRET source 3EG J0241+6103 which also shows evidence for 26.5 day modulation in the GeV band (Massi 2004). More recently, LS I +61$^{\circ}$ 303 has been detected as a variable TeV gamma-ray source (Albert et al. 2006, Acciari et al. 2008) with maximum emission observed near apastron.

 LS I +61$^{\circ}$ 303 is one of only three reliably detected TeV binaries: the other two being LS 5039 (Aharonian et al. 2005a) and PSR B1259-63 (Aharonian et al. 2005b). PSR 1259-63 is a confirmed binary pulsar (Johnston et al. 1992a,b) whereas the nature of both LS 5039 and LS I +61$^{\circ}$ 303 is still under debate. The two main competing scenarios which can explain these systems are $\textit{microquasar}$ (i.e. non-thermal emission powered by accretion and jet ejection) or $\textit{binary pulsar}$ (i.e. non-thermal emission powered by the interaction between the stellar and pulsar winds). The microquasar model used to describe LS I +61$^{\circ}$ 303 is supported by evidence for strong jet outflows (Massi 2001). However, this model suffers from the failure to detect blackbody X-ray spectra expected in an accretion scenario. The microquasar scenario has not been ruled out and is still the subject of much theoretical work, for example, see Romero et al. (2007). The binary pulsar model is most strongly supported by VLBA data (Dhawan 2006) which reveal a cometary radio structure around LS I +61$^{\circ}$ 303 that is interpreted as due to the interaction between the pulsar and Be star wind structures. However, there is currently no detection of pulsed radio or X-ray emission confirming the presence of a pulsar. Possible models for LS I +61$^{\circ}$ 303 will be discussed further in Section 4.

X-ray monitoring campaigns conducted with RXTE (Harrison et al. 2000, Greiner and Rau 2001), ROSAT (Taylor et al. 1996), Chandra (Paredes et al. 2007), Beppo-Sax and XMM-Newton (Sidoli et al. 2006) show that LS I +61$^{\circ}$ 303 is a highly variable hard X-ray source with flux levels modulated with the 26.5 day orbital period, the highest flux usually appearing between orbital phases 0.4 and 0.9. The XMM-Newton observations also detail very fast changes of flux, with fluxes doubling over the span of 1000 seconds (Sidoli et al. 2006). This result of kilosecond scale variability in the X-ray band has also been shown in Esposito et al. (2007) where the authors analyze Swift observations of LS I +61$\circ$ 303 taken in 2006. These 2006 Swift observations are reanalyzed and presented in this work.

Chandra observations (Paredes et al. 2007) detail fast variability of the flux levels, while also showing evidence for extended X-ray emission reaching between 5'' and 12.5" to the north of LS I +61$^{\circ}$ 303. This provides an indication that particle acceleration may be taking place as far away as 0.05-0.12 parsecs from LS I +61$^{\circ}$ 303. Recent RXTE observations (Smith et al. 2009), which cover a total of six orbital cycles, show no strong orbital modulation of the 2-10 keV X-ray flux, but a highly significant correlation between spectral index and flux levels. These observations (which are used in this work) show the presence of three large flares, the largest peaking at a flux value of 7.2 ($^{+0.1}_{-0.2}$) $\times$10$^{-11}$ ergs s$^{-1}$ cm$^{-2}$. Closer examination of these flaring states shows that the X-ray flux from LS I +61$^{\circ}$ 303 doubles within timescales of $<$2 s, indicating that the X-ray emission region is less than 10$^{11}$ cm in extent.

The MAGIC collaboration first detected LS I +61$^{\circ}$ 303 as a variable TeV source above 200 GeV using observations made in 2005/2006 (Albert et al. 2006). This dataset covered six orbital cycles in the phase range $\phi$=0.1$-$0.8 and a strong gamma-ray flux was detected during orbital phases $\phi$=0.4$-$0.7, with the observed flux peaking at 16$\%$ of the Crab Nebula flux at phase $\phi$=0.6. The source was not detected during other orbital phases (i.e. $\phi$=0.1$-$0.3 and $\phi$=0.7$-$0.8), which includes the periastron passage. The extracted photon spectrum from 0.2 to 4 TeV measured by MAGIC is well fit by a power-law with differential spectral index $\alpha=$2.6$\pm$0.4$_{stat+sys}$. 

The MAGIC detection was subsequently confirmed by the VERITAS collaboration which detected the source in $>$300 GeV gamma rays over five orbital phases (Acciari et al. 2008). Overall, the two published TeV detections on this source indicate that it is only active at TeV energies near the apastron passage of the compact object in its orbit around the Be star.  Additional observations conducted by the MAGIC collaboration in 2006 (Albert et al. 2009) sampled a total of four orbital cycles, accruing data in all orbital phases. From these observations, a TeV period of 26.8 ($\pm$0.2) days is derived, consistent with the accepted orbital period of the binary.

Although Albert et al. (2008a) uses VLBA, Swift, and MAGIC TeV data points to claim a weak correlation between TeV and X-ray points, there has not yet been shown to be any statistically significant correlation between the two bands. Most of the favored models predict TeV emission via the inverse-Compton mechanism, which would result in correlated emission in the X-ray band, so it is important to simultaneously measure the flux at TeV and X-ray energies. Dedicated studies at both X-ray and TeV energies are also necessary to understand the variability of this source across the electromagnetic spectrum.

\section{Observations and Analysis}

\subsection{VERITAS Observations}
The VERITAS array (Weekes et al. 2002) of imaging atmospheric Cherenkov telescopes (IACTs) located in southern Arizona (1268 m.a.s.l., 31$^{\circ}$40'30''N, 110$^{\circ}$57'07'' W) began 4-telescope array observations in April 2007 (Maier et al. 2007) and is the most sensitive IACT instrument in the Northern Hemisphere. The array is composed of four 12m diameter telescopes, each with a Davies-Cotton tessellated mirror structure of 345 12m focal length hexagonal mirror facets (total mirror area of 110 m$^{2}$). Each telescope focuses Cherenkov light from particle showers onto its 499 pixel PMT camera. Each pixel has a field of view of 0.15$^{\circ}$, resulting in a camera field of view of 3.5$^{\circ}$. VERITAS has the capability to detect and measure gamma rays in the 100 GeV to 30 TeV energy regime with an energy resolution of 15-20$\%$ and an angular resolution of 0.1$^{\circ}$ on an event by event basis. 

Observations for this work were taken in ``wobble'' mode, where the source is offset from the center of the field of view allowing for simultaneous determination of both the source flux and the background (Fomin et al. 1994). Events passing three levels of hardware trigger criteria\footnote[1]{See Holder et al. (2006) and Maier et al. (2007) for description of the VERITAS hardware trigger layout.} were recorded and candidate gamma-ray events were chosen through selection criteria based upon image quality and shape parameters. Event images were selected based upon their total integrated charge (\textit{size} cut), the image moments (\textit{Mean Scaled Width and Length} cuts, Konopelko et al. 1995) and the reconstructed point of origin within the field of view (\textit{$\theta^{2}$} cut).\footnote[2]{See Acciari et al. (2008) for a detailed description of the VERITAS data reduction and analysis procedures.} For the 2007/2008 dataset, a significant fraction of the data ($\sim$50$\%$) were taken under partially moonlit sky in order to maximize the observing time. For these observations, an increased analysis threshold of 500 GeV was imposed. In order to provide a more accurate comparison between the 2006/2007 and 2007/2008 datasets, the 2006/2007 TeV data presented in Acciari et al. (2008) are presented in this work reanalyzed with a 500 GeV analysis threshold as well. 

\subsection{X-ray Observations with RXTE and Swift}

The two RXTE (Swank 1994) datasets used for this work were accumulated first as a result of a Target of Opportunity (ToO) observation request in October 2006 and then as a campaign of dedicated observations in 2007 and 2008. The October 2006 ToO request resulted in ten pointings (one pointing every other day), and a total exposure time of $\sim$9 ks from 10/13/06 to 10/31/06, spanning an orbital phase range of $\phi = 0.14 - 0.83$. Observations with RXTE in 2007/2008 consist of a 1 ks pointing every other day from 08/28/07 until 02/02/2008. These observations cover six full 26.5 day orbital cycles. The XSPEC 12 software package (Arnaud 1996) was used to fit spectra extracted from all available Proportional Counting Units (PCUs) in each night's observations with a simple absorbed power-law, assuming a fixed absorbing hydrogen column density (N$_{H}$) of 0.75$\times10^{22}$ cm$^{-2}$ (Kalberla et al. 2005). This spectral fit was then integrated over the 2-10 keV range in order to determine a flux for each pointing. All RXTE measurements shown in this work are reported with 1$\sigma$ statistical errors. To produce a single spectrum for multiple pointings (as in the analysis performed in Section 3.3), data from PCU 2 only were used because this was the only PCA unit to remain active for all observations. For additional details of the RXTE analysis performed here, see Smith et al. (2009).

The Swift-XRT observations span the period from September 2006 through September 2007, with a total of 97.3 ks observing time. For a description of Swift and the XRT instrument see Gehrels et al. (2004) and Burrows et al.(2005). This dataset is composed of many $\sim$1 ks pointings which are combined in bins approximately one day wide. The maximum span of a single binned observation is two days.  The Swift-XRT data were screened and processed using the most recent versions of standard Swift tools: Swift Software version 2.8, ftools version 6.5, and XSPEC version 12.4.0. The xrtpipeline task $\textit{xrtmkarf}$ generated the ancilliary response files. The Swift-XRT spectral analysis was made with data extracted in the 0.3-10 keV energy band in ``Photon Counting'' mode. Circular source and background regions with radii of 20'' and 60'', respectively, were used. For spectral analysis, a bin size of 20 cts/bin was generally used; fewer counts per bin were accepted for exposures with less than 150 net counts in the source region. Spectral fits were calculated assuming an absorbed power-law model with the galactic hydrogen column density fixed at 0.75$\times10^{22}$ cm$^{-2}$. Flux values and associated 1 $\sigma$ statistical errors were then calculated by integrating the fitted spectra over the 2-10 keV range.

\section{Results}
\subsection{VERITAS Results}

The TeV dataset used in this work covers 2, 3, and 4-telescope observations made from September 2006 to February 2008, and includes a total of nine 26.5 day orbital cycles of the binary system (see Tables 1 and 2). Figure 1 shows the TeV lightcurve from both years binned by orbital phase. In this figure, excesses with significance above 2$\sigma$ are shown as points with error bars, with all other points being shown as 95$\%$ confidence level (Helene 1983) upper limits. The data from the 2006/2007 observing season (taken with both 2 and 3-telescope arrays) comprised a total of 43.6 hours of observations (after data quality selection) during the orbital phases of $\phi$=0.2$-$0.9 with significant coverage of phases $\phi$=0.4-0.9 (see Acciari et al. 2008). For the entire 2006/2007 dataset, the source was detected at the 8.8$\sigma$ significance level (128 excess events) for emission above 500 GeV. The source was detected as an active TeV source only during apastron phases $\phi=$0.5$-$0.9, with the largest observed fluxes between phases $\phi$=0.6 and $\phi=$0.8 (see Table 1). VERITAS observations made within the same phase range measured flux values ranging from 10-20$\%$ of the Crab Nebula flux (100$\%$ Crab Nebula flux measured as 5.8$\times$10$^{-11}$cm$^{-2}$s$^{-1}$ above 500 GeV). These observed fluxes are similar to those measured by MAGIC between phases $\phi$=0.6 and $\phi$=0.8 (Albert et al. 2006). The differential photon spectrum extracted from the 2006/2007 observations (Acciari et al. 2008) is well fit by a power-law described by dN$_{\gamma}$/dE=(2.39$\pm$0.32$_{stat}\pm$0.6$_{sys}$)$\times$$\frac{E}{1 \space TeV} $$^{-2.4\pm0.2_{stat}\pm0.2_{sys}}$$\times$10$^{-12}$ cm$^{-2}$ s$^{-1}$ TeV$^{-1}$ for emission above 300 GeV, in agreement with the spectrum derived from MAGIC observations. 

The 2007/2008 season dataset is composed of a total of 20.7 hours of four telescope observations taken between October 2007 and January 2008, spanning five separate orbital cycles. From these observations the source was detected at a significance level of 7.3$\sigma$ (71 excess events) for emission above 500 GeV. These observations cover all orbital phases, however, due to factors such as poor weather conditions not all phase bins were covered with equal exposure time (see Figure 1). The source was significantly detected during the orbital phases of $\phi$=0.7-0.9 (see Figure 1 and Table 2) with observations taken during this orbital phase range detecting 41 excess events in 436 minutes, corresponding to an average flux of 4.0$\%$($\pm$0.6$\%$) of the Crab Nebula flux above 500 GeV at a 6.5$\sigma$ statistical significance.

                                                                                                                                                                                                                                                                                                                                                                                                                                                                                  For the 2007/2008 observations, the differential photon spectrum extracted from phases $\phi$=0.5$-$0.8 (the same orbital phases from which the 2006/2007 spectrum was extracted in (Acciari et al. 2008)) is well fitted by a power-law described by dN$_{\gamma}$/dE=(2.43$\pm$0.78$_{stat}\pm$0.6$_{sys}$)$\times$$\frac{E}{1\space TeV}$$^{-2.6\pm0.6_{stat}\pm0.2_{sys}}$$\times$10$^{-12}$ cm$^{-2}$ s$^{-1}$ TeV$^{-1}$, in good agreement with the 2006/2007 spectrum (see Figure 2), although the precision on the spectrum is reduced due to more limited statistics.

To address the question as to whether there might be some low level TeV emission occurring outside of the apastron passage phases in the 2007/2008 observations, we combined the data taken outside phases $\phi$=0.5-0.9. This dataset shows an excess of 28 events in 693 minutes of observation time, which is consistent with an average flux value of 1.7$\%$($\pm$0.6$\%$) of the Crab Nebula flux at a 3.6$\sigma$ statistical significance. However, we find that this excess is principally due to data from phases $\phi$=0.2-0.3 which show an excess of 15 events in 230 minutes of observation time. This excess corresponds to a flux of 2.7$\%$($\pm$0.8$\%$) of the Crab Nebula at a 3.3$\sigma$ statistical significance (post-trial significance of 2.8$\sigma$ for six trials). While we do not consider this result to be statistically significant evidence for TeV emission outside of apastron passage, the excess indicates the possibility of a low level flux which warrants further investigation.

LS I +61$^{\circ}$ 303 has been demonstrated by previous observations (Acciari et al. 2008, Albert et al. 2006, Albert et al. 2008, Albert et al. 2009) to be variable TeV source. To confirm this result we performed a test for the probability of the source having a constant flux over both years during which the dataset presented here was accrued. We tested 200 individual fluxes between 1$\%$ and 20$\%$ Crab flux strength and computed the corresponding probabilities that these constant fluxes would provide a reasonable fit to the observed nightly flux upper limits and detections based on the observed excess events, livetime, and calculated effective area of each night's data. We found no probable constant flux fit to the observed data, with the best fit constant flux corresponding to a 6.3$\%$ Crab Nebula flux above 500 GeV. This constant flux value resulted in a reduced $\chi^{2}$ value of 4.1 (for 55 degrees of freedom), corresponding to a probability of less than 10$^{-16}$ that LS I +61$^{\circ}$ 303 presented a constant flux over the two years of data presented here. 

\subsection{RXTE and Swift Results}

 The X-ray flux in the 2006/2007 season, as seen by both instruments, is highly variable, ranging between approximately 0.5 $\times 10^{-11}$ ergs cm$^{-2}$s$^{-1}$ and 3 $\times 10^{-11}$ ergs cm$^{-2}$s$^{-1}$ over the 26.5 day orbit (see Figures 3 and 4). Similar variability is also present in the 2007/2008 dataset which also shows the presence of three exceptionally large X-ray flares, reaching a peak flux on MJD 54356.96 of 7.2$\times 10^{-11}$ ergs cm$^{-2}$ s$^{-1}$ during a $\sim$500 s integration window (see Figure 4, top panel). This flare is the largest X-ray flare detected from this source, a factor of 2-3 larger than any previously measured. There were several other powerful X-ray flares occurring shortly after this flare in the RXTE data. Further analysis of these flares (Smith et al. 2009) shows that the X-ray flux from LS I +61$^{\circ}$ 303 undergoes doubling over timescales as short as several seconds, as well as changing by up to a factor of six in several hundred seconds. This rapid variability likely explains the apparent disagreement between the RXTE and Swift data points in Figure 4. The fit spectra for both the RXTE and Swift nightly observations are also variable, with indices ranging from 1.4 to 2.6 over the span of the observations. A detailed analysis of the 2007/2008 RXTE data shows that there is a strong correlation between the spectral index and flux values observed from the system (Smith et al. 2009), with the spectrum hardening as luminosity increases.

\subsection{TeV and X-ray Combined}

Given the indication in Albert et al. (2008) of a possible correlation between X-ray and TeV emission, we combined both datasets in order to attempt to look for further evidence of this correlation. For the 2006/2007 observing season there were contemporaneous data for three of the five orbital cycles observed by VERITAS.  While there were several examples of contemporaneous observations which appeared to show similar behavior in both bands,  (see, for example, observations taken between 2006/10/27 and 2006/10/30 in Figure 3) there were also contemporaneous observations which showed elevated flux in the X-ray regime with no corresponding observed TeV flux (see, for example, observations from 2006/11/21 to 2006/11/23). The 2007/2008 combined dataset showed conflicting results as well, with some contemporaneous observations showing similar behavior between the X-ray and TeV bands while other observations did not. For example, the observations taken by VERITAS from 2007/10/28 to 2007/11/05 (see Figure 4) showed a 5.2 $\sigma$ detection of a 6$\%$ Crab Nebula flux between phases 0.7-0.8, whereas only flux upper limits can be placed on emission preceding this (phases $\phi=$0.5-0.6). The X-ray emission is very similar during both of these observations, which calls into question any strong correlation.

Although significant TeV flux detections are relatively sparse throughout both seasons, a test for any correlations present between the X-ray and TeV data is performed. A Z-transformed Discrete Correlation Function (ZDCF) (Alexander 1997) is computed for the X-ray and TeV data from both seasons. This method of correlation testing has been used before for X-ray/TeV datasets in Blazejowski et al. (2005) and was shown to be more effective in finding any present correlations than standard discrete correlation testing on datasets which are sparsely populated (Alexander 1997). There are no statistically significant features present in either test. However, both TeV sets are poorly sampled and it is not clear that this lack of correlation is intrinsic to the source, or due to sparse data sampling.

To examine whether or not the sampling presented here would be sufficient to detect a correlation, if one exists, we generated two continuous lightcurves with similar properties to each year's X-ray observations. Each of these simulated lightcurves was then duplicated and sampled at the times corresponding to the real X-ray and TeV data for each year, with the errors on the fluxes corresponding to the real errors on the TeV and X-ray data. Both sets of lightcurves were then tested using the ZDCF test with no correlations resulting. This demonstrates that even if the TeV and X-ray emission were perfectly correlated, the sampling provided by the observations detailed in this work would not result in a correlation using the methods described above. Given that this correlation would not be apparent under even the best case scenario with the TeV and X-ray sampling provided in this work, any claim of correlated emission between the two bands cannot be justified with the currently available data.

\section{Interpretation and Conclusions}

The data presented in this work show that the X-ray and TeV emission from LS I +61$^{\circ}$ 303 is highly variable over the entire accrued dataset from 2006 to 2008. A gamma-ray signal was significantly detected over the 2006/2007 observing season during apastron passage at a peak flux level of 15-20$\%$ Crab Nebula flux (above 500 GeV), and during the 2007/2008 observations at a $<$5$\%$ Crab Nebula flux during apastron passage (above 500 GeV). Although LS I +61$^{\circ}$ 303 has only been detected during orbital phases $\phi$=0.5-0.9 (including MAGIC observations), there is marginal evidence for emission outside of apastron passage as suggested by both the 2007/2008 VERITAS observations as well as the observations conducted by MAGIC (Albert et al. 2009). Given the relatively sparse TeV data detailed in this work, we are unable to place a constraint on periodicity within the TeV signal as reported in Albert et al. (2008).

The X-ray flux from LS I +61$^{\circ}$ 303 is also variable, with strong outbursts occurring at multiple regions of the orbit. The ZDCF analysis of the quasi-contemporaneous X-ray and TeV dataset does not show evidence for a correlation between the two bands. However, due to the lack of dense TeV coverage overlapping with X-ray observations, our sensitivity to such a correlation is inadequate. More specifically, observations conducted with RXTE, Swift, and XMM-Newton (Smith et al. 2009, Paredes et al. 2007, Sidoli et al. 2006) show that  the X-ray flux from LS I +61$^{\circ}$ 303 can change significantly over short timescales (up to a factor of 6 over several hundred seconds). If the X-ray and TeV emission are indeed correlated on fast timescales, truly simultaneous coverage in both bands would be necessary to confirm this correlation. 

Although the data presented here do not conclusively rule out or reinforce any of the proposed models (i.e. binary pulsar or microquasar), the derived TeV and X-ray spectra can be compared to recent model predictions from both scenarios.  Since both TeV spectra presented here are composed of data taken over several orbital cycles, it is not possible to construct a truly simultaneous or contemporaneous spectral energy distribution for the data examined in this paper. Instead, the RXTE data which fell between orbital phases 0.5 and 0.8 (the phases from which the TeV spectrum was derived in both seasons) were integrated into a single spectrum and were fit by the same procedure as described above for the nightly RXTE points. This resulted in an X-ray spectrum from 3 to 10 keV, which is well fit by the power-law 5.84 ($\pm$0.06)$\times$10$^{-3}$$\times$E$^{-1.89\pm0.05}$ cm$^{-2}$ s$^{-1}$ keV$^{-1}$. This spectrum is plotted along with the EGRET spectrum from Hartmann et al. (1999), and the VERITAS TeV spectra from Acciari et al. (2008) in Figure 8.

An example of a binary pulsar model with a broadband SED prediction is the compactified pulsar wind scenario of Zdziarski, Neronov and Chernyakova (2008) (ZNC). In this model, the X-ray to TeV emission is powered by inverse Compton scattering off stellar UV photons by energetic electrons injected from the pulsar wind into the fast, clumpy polar wind of the Be star, as well as by the more traditional binary pulsar emission mechanism of a shock front located at the interaction point of the Be equatorial and pulsar winds. This model also predicts observable variations in the high-energy emission along the orbit due to accompanying density variations in the Be equatorial wind, as well as inhomogeneities introduced by the mixing of the fast polar and pulsar winds. The model prediction plotted in Figure 8 is a three part broken power-law (similar to that in Chernyakova et al. (2006)) which represents the high X-ray state of the system. The model performs well in predicting the observed spectral energy distribution, with the resulting TeV spectrum prediction of index -2.76 falling within the allowed error range of the VERITAS fit (-2.40$\pm$0.39$_{stat+sys}$). We note that in binary pulsar models the dominant emission mechanism is dictated by the magnetic field strength at the shock, which is in turn dictated by the so called ``stand-off'' distance, or the distance from the compact object to the shock front. When the stand-off distance is smallest (i.e. when the stellar wind strength is greatest) near periastron, the magnetic field strength is greatest, allowing the synchrotron loss channel into hard X-rays to dominate, quenching the production of TeV gamma rays via the inverse-Compton process. The possible existence of TeV emission near periastron passage calls into question the validity of this prediction in the binary pulsar scenario. If confirmed with further observations, the appearance of periastron TeV emission would necessitate modifications to this model.

 An example of a microquasar model which may offer an explanation for the observed emission is that of Gupta and Boettcher (2006). This scenario provides a time-dependent leptonic jet framework that models synchrotron, synchrotron self-Compton, and inverse-Compton (on stellar UV photons) losses resulting from an accretion powered jet within the system. While all mechanisms contribute in varying amounts, the synchrotron contribution dominates the X-ray emission, while the inverse Compton contribution (both external Compton and synchrotron self-Compton) is only significant above MeV energies. (see the double humped structure in Figure 8). This model examines the hypothesis that the observed TeV variability may be able to be interpreted solely as a geometrical effect arising from absorption in the dense photon field of the star. The regions of highest TeV production, therefore, are those that have limited exposure to this dense absorption field such as the $\phi$=0.5 orbital region which is plotted in Figure 8.  As can be seen, while the synchrotron contribution adequately reproduces the observed X-ray spectrum, the inverse Compton contributions underproduce both the EGRET and VERITAS spectra, predicting a cutoff at a few TeV. This fit could most likely be improved, however, if the constraint of absorption being the only contributing factor to the TeV variability was removed. Given the possible existence of TeV emission outside of apastron passage, this would seem to be a necessary modification. 

Recently, Swift-BAT has reported the detection of a short (0.23 s), extremely powerful X-ray burst with a luminosity of 10$^{37}$ erg s$^{-1}$ in the 15-150 keV energy range within the 90$\%$ containment radius of LS I +61$^{\circ}$ 303 (Barthelmy et al. 2008).  While Barthelmy et al. (2008) notes the possibility that this emission episode was due to an unrelated short gamma-ray burst, they claim that the evidence is in favor of activity from a source within LS I +61$^{\circ}$ 303 (Barthelmy et al. 2008). Further analysis of Barthelmy et al. (2008) by Dubus and Giebels (2008) interprets this burst as evidence for magnetar activity within LS I +61$^{\circ}$ 303, the first within a high mass X-ray binary. If confirmed by subsequent observations, this type of extremely powerful bursting may help to resolve the question of the identification of LS I +61$^{\circ}$ 303. 

In conclusion, the multiwavelength observations reported here demonstrate clear variability of LS I +61$^{\circ}$ 303 in the high energy regime, however, the TeV data has insufficient sampling to constrain the correlation between the X-ray and TeV bands. Additionally, while neither conclusively ruling out nor confirming either the microquasar or binary pulsar scenarios, the observations reported here show marginal evidence for TeV emission near periastron passage, a feature which, if confirmed, may be necessary to incorporate into future models.  Future simultaneous, multiwavelength observations with instruments such as VERITAS and MAGIC in the TeV regime, combined with GeV observations by the Fermi Gamma-ray Space Telescope, and X-ray monitoring by various instruments will aid in the deeper understanding of this unpredictable and exciting source.

\acknowledgments

This research is supported by grants from the U.S. Department of Energy, the U.S. National Science Foundation,
the Smithsonian Institution, by NSERC in Canada, by Science Foundation Ireland and by PPARC in the U.K..
We acknowledge the excellent work of the technical support staff at the FLWO and 
the collaborating institutions in the construction and operation of the instrument.

The submitted manuscript has been created by UChicago Argonne, LLC,             
Operator of Argonne National Laboratory (``Argonne''). Argonne, a U.S.            
Department of Energy Office of Science laboratory, is operated under            
Contract No. DE-AC02-06CH11357. The U.S. Government retains for itself,         
and others acting on its behalf, a paid-up nonexclusive, irrevocable            
worldwide license in said article to reproduce, prepare derivative              
works, distribute copies to the public, and perform publicly and display        
publicly, by or on behalf of the Government.

\begin{figure}
\begin{center}
   \includegraphics[width=\textwidth,height=120mm]{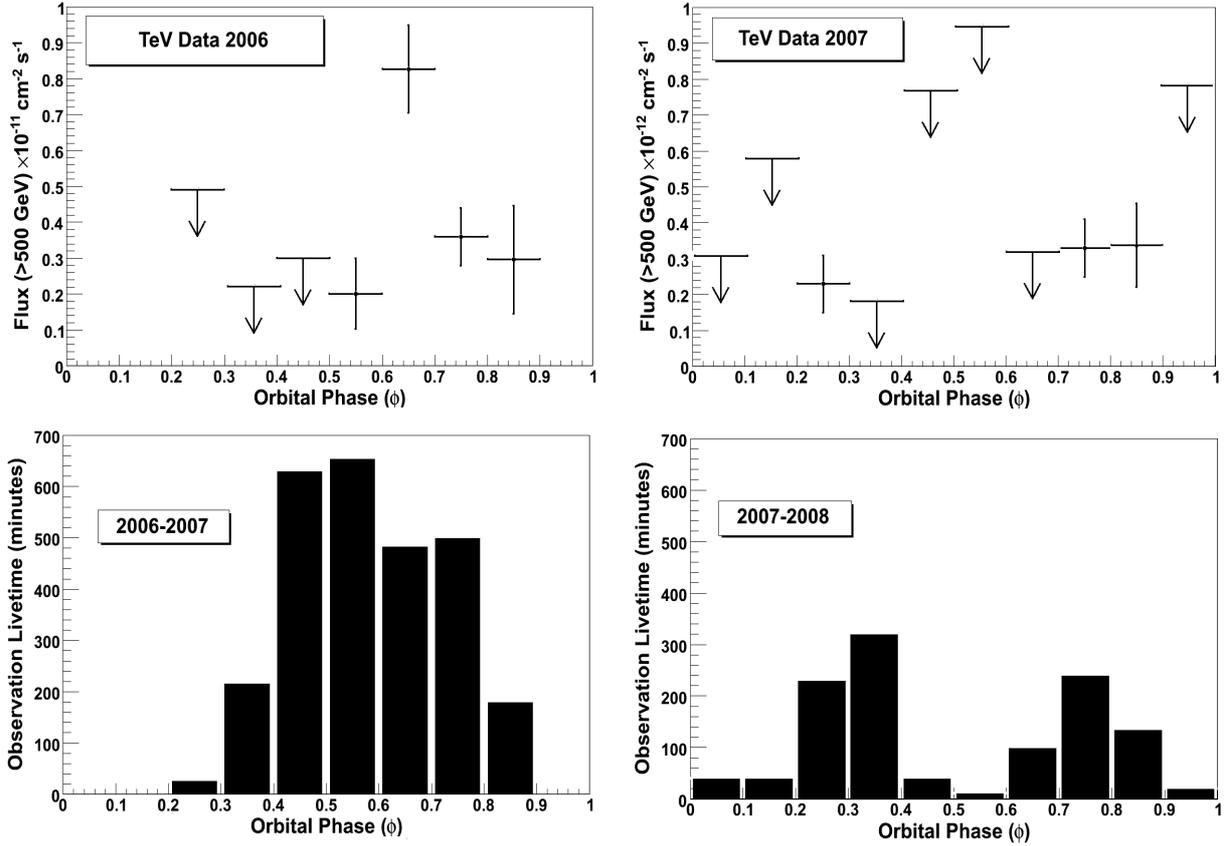}
\end{center}
\caption[]{The results of the 2006/2007 (top left) and 2007/2008 (top right) VERITAS TeV observations of LS I +61$^{\circ}$ 303.  Points with error bars represent signal detections above a 2$\sigma$ significance threshold, points with arrows represent 95$\%$ confidence flux upper limit points. The bottom two panels show exposure times (dead time corrected) per orbital phase bin for each observation season.}
\end{figure}

\begin{figure}
\begin{center}
   \includegraphics[width=0.6\textwidth,height=90mm]{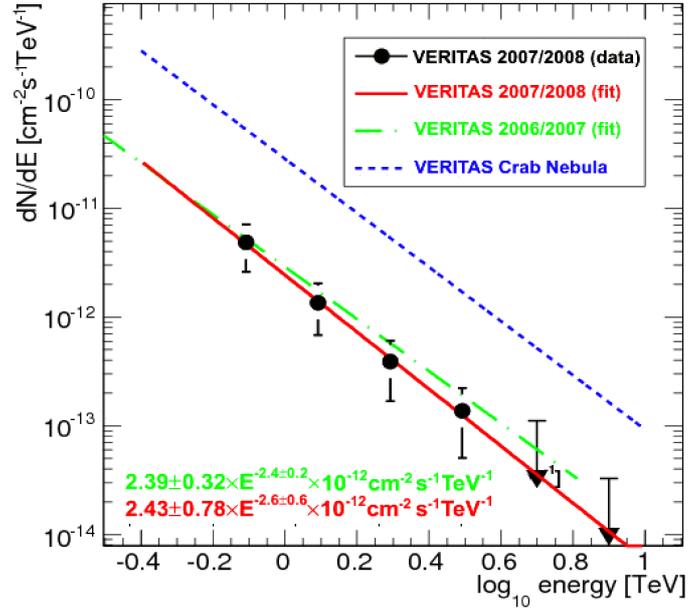}
\end{center}
\caption[]{The differential photon spectra of LS I +61$^{\circ}$ 303 as measured by VERITAS during orbital phases 0.5$-$0.8 from both the 2006/7 (Acciari et al. 2008) and 2007/8 seasons, along with fitted Crab Nebula spectrum as measured by VERITAS. See section 3.1 for a description of the data and power-law fits.}
\end{figure}


\begin{figure}
\begin{center}
   \includegraphics[width=\textwidth,height=180mm]{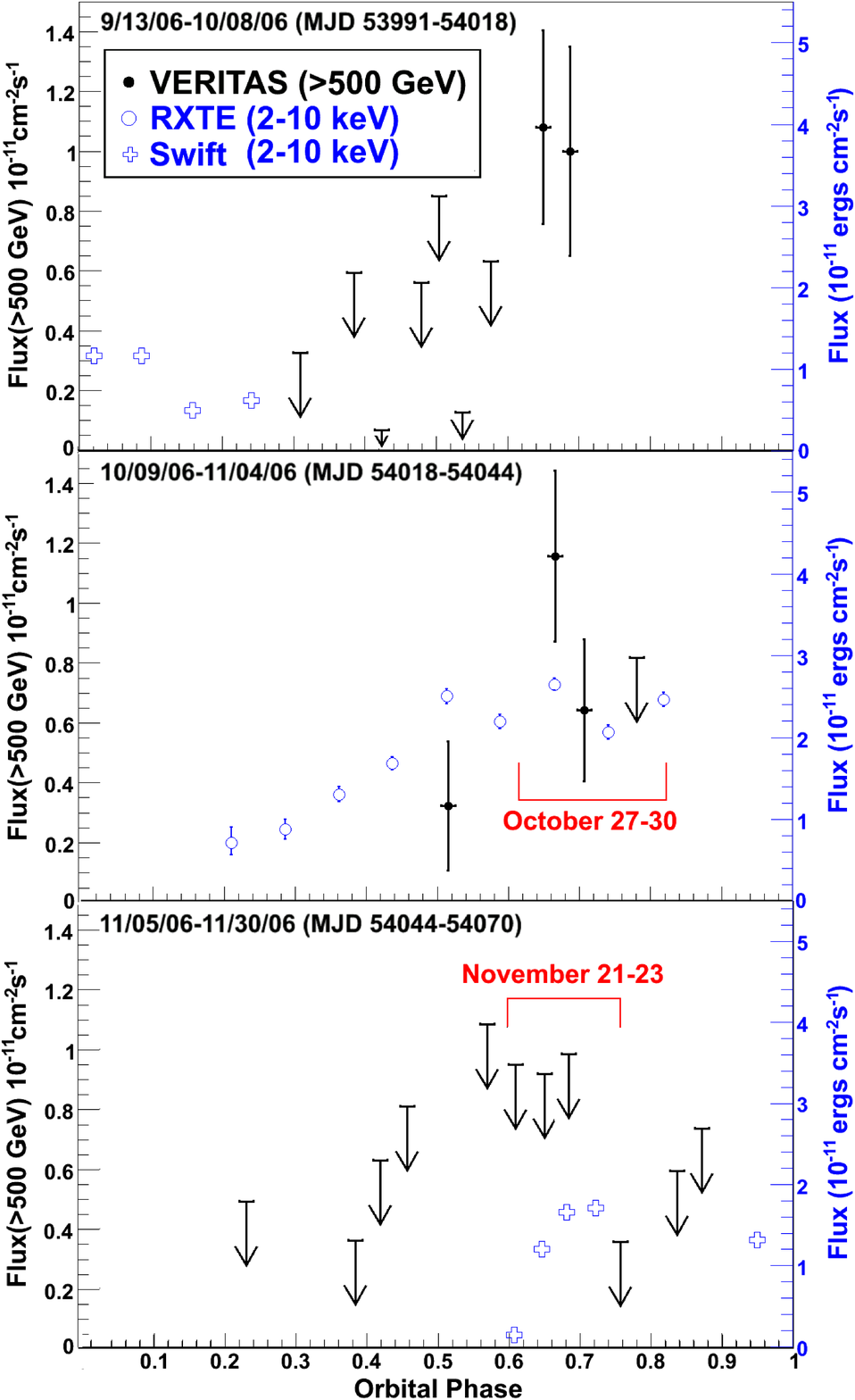}
\end{center}
\caption[]{The results of the 2006/2007 TeV and X-ray observations. The vertical scale on the right (blue) represents the X-ray flux only, with the left vertical scale representing TeV flux.}
\end{figure}
\begin{figure}
\begin{center}
   \includegraphics[width=\textwidth,height=230mm]{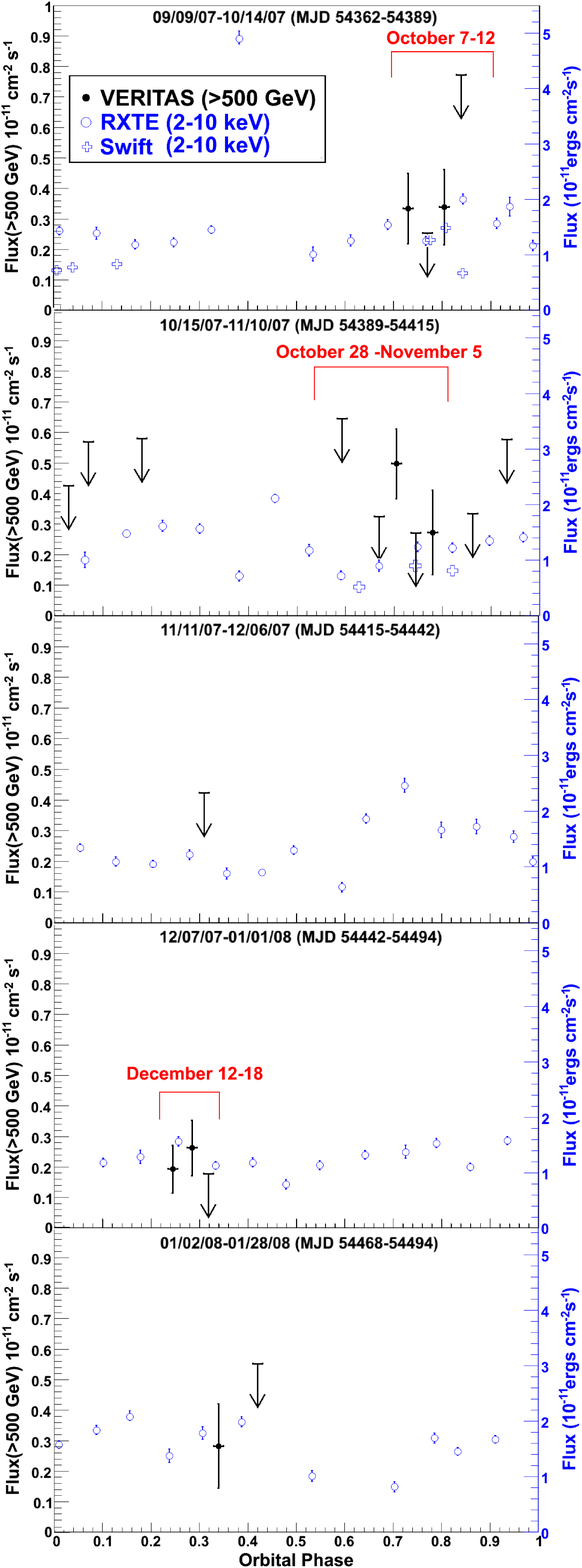}
\end{center}
\caption[]{The results of the 2007/2008 TeV and X-ray observations. Scales are as in figure 4. }
\end{figure}



\begin{figure}
\begin{center}
   \includegraphics[width=\textwidth,height=100mm]{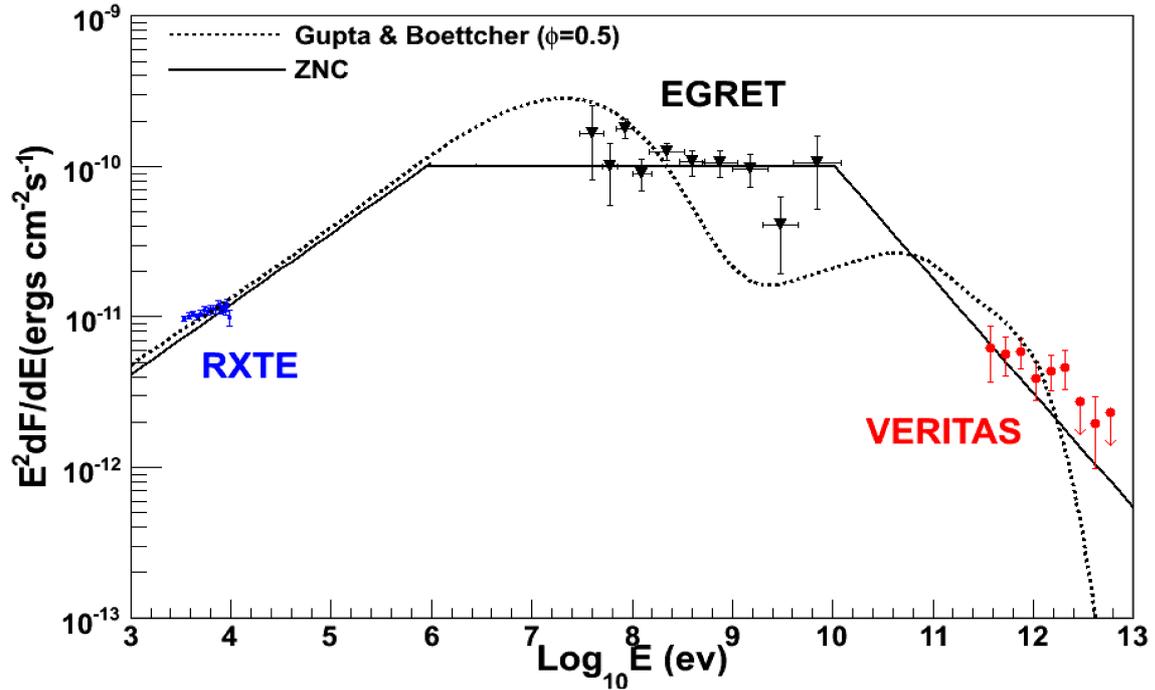}
\end{center}
\caption[]{The spectral energy distribution of LS I +61$^{\circ}$ 303 compared to the models of Gupta and Boettcher (2006) shown by a dashed line and ZNC (2008) shown by a solid line. The Gupta and Boettcher spectrum is the model's prediction at orbital phase 0.5, whereas the ZNC spectrum is that model's prediction at a general high emission state which is not defined in terms of orbital position. }
\end{figure}

\pagebreak
\pagebreak

\begin{table}
  \centering
\caption{VERITAS TeV Observations of LS I +61$^{\circ}$ 303 2006/2007}
\renewcommand{\arraystretch}{0.6}
\begin{tabular}{|>{\small}c|>{\small}c|>{\small}c|>{\small}c|}
\hline
$\textbf{Date Observed}$ & $\textbf{Orbital Phase ($\phi$)}$&$\textbf{ Significance ($\sigma$)}$&$\textbf{Flux($>$500 GeV)}$ \\
$\textbf{MJD}$& & &$\textbf{$\times$10$^{-11}$ cm$^{-2}$ s$^{-1}$}$\\ \hline
53999.4 &  0.31  & -0.5 & $<$0.33         \\ \hline
54001.3 &  0.382 & 0.2 &  $<$0.59         \\ \hline
54002.4 &  0.424 & 0.0 &  $<$0.68         \\ \hline
54003.4 &  0.477 & 1.04 & $<$0.56          \\ \hline
54004.3 &  0.5   & 0.74 & $<$0.85          \\ \hline
54005.4 &  0.537 & -1.49 &$<$0.13          \\ \hline
54006.4 &  0.575 & 1.14 & $<$0.63          \\ \hline
54008.4 &  0.65  & 4.39 &  1.08$\pm$0.32   \\ \hline
54009.4 &  0.688 & 3.81 &  1.00$\pm$0.35    \\ \hline
54031.3 &  0.515 & 2.04 &  0.32$\pm$0.22    \\ \hline
54035.4 &  0.666 & 5.4 &   1.15$\pm$0.29    \\ \hline
54036.4 &  0.707 & 3.62 &  0.64$\pm$0.24    \\ \hline
54038.4 &  0.782 & 1.57 & $<$0.81           \\ \hline
54050.2 &  0.23  & -0.32 &$<$0.49           \\ \hline
54054.3 &  0.383 & 0.44 & $<$0.36          \\ \hline
54055.3 &  0.42  & 1.14 & $<$0.63           \\ \hline
54056.3 &  0.458 & 1.17 & $<$0.81           \\ \hline
54059.3 &  0.57  & 1.86 & $<$1.09         \\ \hline
54060.3 &  0.61  & 2.86 & 0.95$\pm$0.46      \\ \hline
54061.3 &  0.65  & 1.21 & $<$0.92          \\ \hline
54062.3 &  0.685 & 0.84 & $<$0.99         \\ \hline
54064.3 &  0.756 & 0.31 & $<$0.35         \\ \hline
54066.3 &  0.836 & 1.99 & $<$0.59         \\ \hline
54067.3 &  0.87  & 1.36 & $<$0.73         \\ \hline
54108.2 &  0.42  & -0.06 &$<$0.22             \\ \hline
54109.2 &  0.45  & 2.15 & 0.32$\pm$0.29     \\ \hline
54110.2 &  0.49  & 0.34 & $<$0.29           \\ \hline
54115.1 &  0.677 & 3.37 & 0.71$\pm$0.28    \\ \hline
54116.1 &  0.715 & 2.97 & 0.39$\pm$0.16    \\ \hline
54117.2 &  0.75  & 2.24 & 0.53$\pm$0.3    \\ \hline
54137.1 &  0.756 & -0.84 &$<$0.38           \\ \hline
54138.1 &  0.793 & 0.59 & $<$0.52          \\ \hline
54139.1 &  0.58  & 0.46 & $<$0.62          \\ \hline
54140.1 &  0.62  & 0.46 & $<$0.47           \\ \hline
54144.1 &  0.77  & 2.24 & 0.71$\pm$0.4    \\ \hline
54147.1 &  0.89  & 0.27 & $<$0.48         \\ \hline

\end{tabular}
\end{table}

\begin{table}
  \centering
\caption{VERITAS TeV Observations of LS I +61$^{\circ}$ 303 2007/2008}
\renewcommand{\arraystretch}{0.6}
\begin{tabular}{|>{\small}c|>{\small}c|>{\small}c|>{\small}c|}
\hline

$\textbf{Dates Observed}$ & $\textbf{Orbital Phase ($\phi$)}$&$\textbf{ Significance ($\sigma$)}$&$\textbf{Flux($>$500 GeV)}$ \\
$\textbf{MJD}$& & &$\textbf{$\times$10$^{-11}$ cm$^{-2}$ s$^{-1}$}$\\ \hline

54381.9 &  0.73 &2.9 &  0.33$\pm$ 0.11\\ \hline
54382.9 &  0.77 &0.02 & $<$0.25        \\ \hline
54383.9 &  0.81 &2.7 &  0.33$\pm$0.12 \\ \hline
54384.9 &  0.84 &1.9 & $<$0.77        \\ \hline
54389.9 &  0.03 &0.32 & $<$0.42        \\ \hline
54390.9 &  0.07 &1.1 & $<$0.56        \\ \hline
54393.8 &  0.18 &1.1 & $<$0.58        \\ \hline
54404.2 &  0.59 &0.76 & $<$0.64        \\ \hline
54406.7 &  0.67 &1.6 & $<$0.32        \\ \hline
54407.8 &  0.71 &4.4 &  0.49$\pm$0.11 \\ \hline
54408.8 &  0.75 &0.63 &  $<$0.27       \\ \hline
54409.7 &  0.78 &2 &  0.27$\pm$0.11 \\ \hline
54411.9 &  0.86 &-0.06 &  $<$0.33       \\ \hline
54413.8 &  0.93 &0.0 &  $<$0.57       \\ \hline
54423.8 &  0.31 &0.11 &  $<$0.42       \\ \hline
54448.6 &  0.25 &2.5 &  0.19$\pm$0.08 \\ \hline
54449.7 &  0.29 &2.9 &  0.26$\pm$0.09 \\ \hline
54450.7 &  0.32 &1.9 &  $<$0.18       \\ \hline
54477.6 &  0.34 &2.1 &  0.28$\pm$1.38 \\ \hline
54479.6 &  0.42 &0.52 &  $<$0.55       \\ \hline

\end{tabular}
\end{table}

\end{document}